\documentclass[conference,compsoc]{IEEEtran}

\usepackage[utf8]{inputenc}
\usepackage{xcolor}
\usepackage{comment}
\usepackage{graphicx}
\PassOptionsToPackage{hyphens}{url}\usepackage[hidelinks]{hyperref}
\usepackage{amssymb}
\usepackage{pifont}
\usepackage{soul}
\usepackage[acronym]{glossaries}
\usepackage{cite}

\glsdisablehyper  
\newacronym{ZTA}{ZTA}{Zero Trust Architecture}
\newacronym{ZT}{ZT}{Zero Trust}
\newacronym{ZTE}{ZTE}{Zero Trust Execution}
\newacronym{TEE}{TEE}{Trusted Execution Environment}
\newacronym{SEV}{SEV}{Secure Encrypted Virtualization}
\newacronym{SME}{SME}{Secure Memory Encryption}
\newacronym{SP}{SP}{Secure Processor}
\newacronym{ES}{ES}{Encrypted State}
\newacronym{SNP}{SNP}{Secure Nested Paging}
\newacronym{SGX}{SGX}{Software Guard Extensions}
\newacronym{TDX}{TDX}{Trust Domain Extensions}
\newacronym{TME}{TME}{Total Memory Encryption}
\newacronym{PEF}{PEF}{Protected Execution Facility}
\newacronym{CCA}{CCA}{Confidential Compute Architecture}
\newacronym{SBA}{SBA}{Service Based Architecture}
\newacronym{SBI}{SBI}{Service Based Interface}
\newacronym{NF}{NF}{Network Function}
\newacronym{NS}{NS}{Network Slice}
\newacronym{NFV}{NFV}{Network Function Virtualization}
\newacronym{NFVI}{NFVI}{Network Function Virtualization Infrastructure}
\newacronym{VNF}{VNF}{Virtualized Network Function}
\newacronym{B5G}{B5G}{Beyond-5G}
\newacronym{MNO}{MNO}{Mobile Network Operator}
\newacronym{SDN}{SDN}{Software Defined Networks}
\newacronym{AMF}{AMF}{Access and Mobility Management Function}
\newacronym{SMF}{SMF}{Session Management Function}
\newacronym{AUSF}{AUSF}{Authentication Server Function}
\newacronym{UPF}{UPF}{User Plane Function}
\newacronym{NEF}{NEF}{Network Exposure Function}
\newacronym{NRF}{NRF}{Network Function Repository Function}
\newacronym{NSSF}{NSSF}{Network Slice Selection Function}
\newacronym{PCF}{PCF}{Policy Control Function}
\newacronym{UDM}{UDM}{Unified Data Management}
\newacronym{SEAF}{SEAF}{Security Anchor Function}
\newacronym{UE}{UE}{User Equipment}
\newacronym{CN}{CN}{Core Network}
\newacronym{5GC}{5GC}{5G Core}
\newacronym{DN}{DN}{Data Network}
\newacronym{RAN}{RAN}{Radio Access Network}
\newacronym{VM}{VM}{Virtual Machine}
\newacronym{SVM}{SVM}{Secure Virtual Machine}
\newacronym{SUCI}{SUCI}{Subscription Concealed Identifier}
\newacronym{SUPI}{SUPI}{Subscription Permanent Identifier}
\newacronym{MANO}{MANO}{Management and Network Orchestration}
\newacronym{OSS/BSS}{OSS/BSS}{Operations Support Systems/Business Support Systems}
\newacronym{BSS}{BSS}{Business Support Systems}
\newacronym{OSS}{OSS}{Operations Support Systems}
\newacronym{FDE}{FDE}{Full Disk Encryption}
\newacronym{TCB}{TCB}{Trusted Computing Base}

\AtBeginDocument{%
  \providecommand\BibTeX{{%
    \normalfont B\kern-0.5em{\scshape i\kern-0.25em b}\kern-0.8em\TeX}}}

\def\BibTeX{{\rm B\kern-.05em{\sc i\kern-.025em b}\kern-.08em
    T\kern-.1667em\lower.7ex\hbox{E}\kern-.125emX}}

\begin{document}

\title{Establishing Trust in the Beyond-5G Core Network using Trusted Execution Environments}

\author{
\IEEEauthorblockN{Marinos Vomvas}
\IEEEauthorblockA{\textit{Northeastern University} \\
Boston, USA \\
vomvas.m@northeastern.edu}
\and
\IEEEauthorblockN{Norbert Ludant}
\IEEEauthorblockA{\textit{Northeastern University} \\
Boston, USA \\
ludant.n@northeastern.edu}
\and
\IEEEauthorblockN{Guevara Noubir}
\IEEEauthorblockA{\textit{Northeastern University} \\
Boston, USA \\
g.noubir@northeastern.edu}
}

\maketitle

\begin{abstract}
 The fifth generation (5G) of cellular networks starts a paradigm shift from the traditional monolithic system design to a Service Based Architecture, that fits modern performance requirements and scales efficiently to new services. This paradigm will be the foundation of future cellular core networks beyond 5G. The new architecture splits network functionalities into smaller logical entities that can be disaggregated logically, physically, and geographically. This affords interoperability between the mobile network operators and commercial software and hardware vendors or cloud providers. By making use of commodity services and products, this system construct inherits the vulnerabilities in those underlying technologies, thereby increasing its attack surface and requiring a rigorous security analysis. In this work, we review the security implications introduced in B5G networks, and the security mechanisms that are supported by the 5G standard. We emphasize on the support of Zero Trust Architecture in 5G and its relevance in decentralized deployments. We revisit the definition of \textit{trust} in modern enterprise network operations and identify important Zero Trust properties that are weakened by the nature of cloud deployments. To that end, we propose a vertical extension of Zero Trust, namely, Zero Trust Execution, to model untrusted execution environments, and we provide an analysis on how to establish trust in Beyond-5G network architectures using Trusted Execution Environments. Our analysis shows how our model architecture handles the increased attack surface and reinforces the Zero Trust Architecture principles in the 5G Core, without any changes to the 5G standard. Finally, we provide experimental results over a 5G testbed using Open5GS and UERANSIM that demonstrate minimal performance overhead, and a monetary cost evaluation.
\end{abstract}

\section{Introduction}

\IEEEPARstart{C}{ellular} networks are constantly evolving, and constitute the cornerstone of global communications for social, entertainment, and critical operation purposes alike.
Thanks to technological advancements, every network generation provides significant quality of service improvements, new features, and improved security and privacy mechanisms.
The upcoming generations promise to revolutionize the way we communicate and access services, within society or business, industry or even medical care. Their use cases include smart factories for Industry 4.0~\cite{5gsmartfactory,industry4}, monitoring devices for Health IoT~\cite{MIoT_monitoring}, autonomous-vehicle communications~\cite{self_driving_cars}, and holographic communications.
Such disparate use cases require drastically different performance guarantees, e.g. high throughput, low latency or ultra-high reliability.
To address this challenge, 5G introduces a major architectural change, redesigning the mobile core.
This change constitutes the foundation for \gls{B5G} networks, such as the 6G and 7G, which mainly improve on the \gls{RAN}~\cite{5g_vs_6g, future_mobile_technologies, 7g_insights, what_about_7g, feature_of_7g}.

\begin{table*}[ht]
\caption{Glossary of acronyms.}
\label{tab:acronyms}
\begin{tabular}{l l l l}
5GC & 5G Core Network   &   SEAF & Security Anchor Function \\
AMF & Access and Mobility Management Function   &   SEV & Secure Encrypted Virtualization   \\
ARM CCA & ARM Confidential Compute Architecture    &    SEV-ES & SEV-Encrypted State    \\
AUSF & Authentication Server Function   &   SEV-SNP & SEV-Secure Nested Paging  \\
B5G & Beyond-5G    &    SGX & Software Guard Extensions \\
CN & Core Network   &   SME & Secure Memory Encryption  \\
DN & Data Network   &   SMF & Session Management Function   \\
FDE & Full Disk Encryption   &  SP & Secure Processor   \\
IBM PEF & IBM Protected Execution Facility   &  SUCI & Subscription Concealed Identifier    \\
MANO & Management and Network Orchestration   & SUPI & Subscription Permanent Identifier    \\
MNO & Mobile Network Operator   &   SVM & Secure Virtual Machine    \\
NEF & Network Exposure Function   & TCB & Trusted Computing Base    \\
NF & Network Function   &   TDX & Trust Domain Extensions   \\
NFV & Network Function Virtualization   &   TEE & Trusted Execution Environment \\
NFVI & Network Function Virtualization Infrastructure   &   TME & Total Memory Encryption   \\
NRF & Network Function Repository Function   &  UDM & Unified Data Management   \\
NS & Network Slice   &  UE & User Equipment \\
NSSF & Network Slice Selection Function   & UPF & User Plane Function   \\
OSS/BSS & Operations Support Systems/Business Support Systems   &   VM & Virtual Machine    \\
PCF & Policy Control Function   &   VNF & Virtualized Network Function  \\
RAN & Radio Access Network   &  ZT & Zero Trust \\
SBA & Service Based Architecture    &   ZTA & Zero Trust Architecture   \\
SBI & Service Based Interface   &   ZTE & Zero Trust Execution  \\
SDN & Software Defined Networks   \\
\end{tabular}
\end{table*}

The \gls{5GC} adapts to diverse use cases by substantially increasing the flexibility of the network.
To accomplish that, \gls{5GC} transitions from the traditional, physical-entity based cellular architecture to a \gls{SBA}. This architecture splits the functionalities previously provided by physical entities into multiple virtual \glspl{NF}, and deploys them in a highly modular infrastructure, where each NF is now deployed as software running on decentralized hardware. Such an adaptive and flexible architecture can be efficiently deployed thanks to recent advances in technologies such as \gls{SDN} and \gls{NFV}~\cite{SDNNFV5G}.

The advantages of an \gls{SBA} are manifold. First, the \gls{CN} can be deployed in various configurations, providing flexibility and cost-efficiency from an operational perspective~\cite{TS_rel16_SON}. Secondly, it simplifies the seamless integration of vendor \glspl{NF} into the architecture using standardized APIs. Moreover, it enables the configuration of logical networks running on top of the physical infrastructure, a paradigm termed Network Slicing~\cite{netslicing_challenges}. This new approach effectively allows on-demand delivery of services like Ultra-Reliable Low-Latency Communications (URLLC), massive Machine Type Communications (mMTC), or enhanced Mobile Broadband (eMBB), by a common infrastructure.
Last but not least, an \gls{SBA} enables the \gls{5GC} to serve as the backbone for the computationally demanding features expected in 6G and 7G.

\textit{Although the virtualization-based paradigm creates a vast range of opportunities, it also expands the attack surface of modern cellular networks.}
It forces the decentralized deployment of network components, on a variety of platforms that may not be owned or managed by the operator.
Such modern heterogeneous systems resolve to \gls{ZT} approaches to complement the traditional perimeter-based security defenses.

In this paper, we extend the notion of \gls{ZT} and define \gls{ZTE} that models untrusted execution environments. We examine how trust in \gls{B5G} cellular networks is reinforced using \glspl{TEE} in the \gls{CN} deployment, such that it adheres to the \gls{ZTE} principles. Finally, we benchmark the performance of our design and find that secure computation can be achieved with minimal overhead, acceptable for the control functionalities of the \gls{5GC}, and at a reasonable monetary cost.

The rest of the paper is structured as follows: In section~\ref{sec:5g_background} we provide background on modern cellular networks, including the revised core architecture in 5G and the path toward 6G and 7G. In section~\ref{sec:sec_model} we describe the security challenges, provide
background on \gls{ZT} pillars and define our extended notion of vertical \gls{ZT}. In section~\ref{sec:tees} we provide background on \glspl{TEE} and
a comparison of the state of the art. Section~\ref{sec:system_design} describes our proposed model architecture along with a detailed security analysis, followed by our experimental results in section~\ref{sec:results}, and our conclusions in section~\ref{sec:conclusion}. We also provide a glossary table with the acronyms used throughout the paper in Table~\ref{tab:acronyms}.

\section{Modern Cellular Networks}
\label{sec:5g_background}
\noindent The latest cellular standard, 5G, introduces a major architectural shift in the \gls{CN}, which is the foundation for \gls{B5G} networks.
At the time of writing, 4G has been deployed at the biggest part of the globe and 5G coverage is rapidly expanding in standalone (5G SA) or non-standalone (5G NSA) configurations~\cite{ericsson_mobility_reports}.
For the foreseeable future, these technologies will likely coexist, each one serving a different purpose.
In the future, B5G networks will focus on advanced connectivity and higher spectrum frequencies, making them ideal for AI-enhanced cellular services and bandwidth-intensive workloads.

Naturally, there are no extensive details or standardization for 6G and 7G beyond technology blogs and experts' discussions, while 8G features are based on speculations. The sixth generation is predicted to unroll by 2030, and its standardization project to complete sometime in 2025~\cite{ericsson_hexa_6g_standard}. Expert discussions for both 6G and 7G suggest that these generations will leverage the \gls{CN} architecture introduced by 5G, and improve on the \gls{RAN} by adding new features and technologies, as well as expanding the use of virtualization to the \gls{RAN}~\cite{what_is_6g, 7g_comms}. On the other hand, 5G has been largely standardized and there are multiple closed and open source software for both the \gls{CN} and \gls{RAN} that support research experiments~\cite{TS_rel17_7, Open5GS, ueransim, oai_cn, oai_ran}. In this section we provide a detailed background on the new architecture of the \gls{5GC} and continue on to describe B5G networks and their relationship with 5G networks.

\subsection{5G Core Network System: Network Functions}\label{sec:5gnfs}
The 5G system architecture is modernized to provide flexibility and versatility, required by the new services envisioned in next-generation cellular communications.
Previous generations implement network functionalities using physical network entities, which are rigid and do not scale easily. Instead, 5G rearranges these functionalities into multiple virtual entities
that can be deployed using container technologies, such as Kubernetes or AWS ECS, and cloud-based platforms.
As an example, the 4G Mobility Management Entity (MME), which handles user mobility, session management, and user authentication,
is decomposed in separate 5G \glspl{NF} that offer equivalent services. The \gls{AMF} handles mobility management, the \gls{SMF} manages the \gls{UE} session, and the \gls{AUSF} provides authentication functions for \gls{UE} authentication.

This split provides scalability, operational flexibility, and the ability to create new services in an agile manner, by simply deploying additional functions in a cloud-native infrastructure.
The 5G system architecture is standardized by the 3GPP organization, and the key network functions in the 5G system as of 3GPP Release 17 are depicted in Figure~\ref{fig:5g_system_architecture}. The system architecture includes 5G \gls{CN}, \gls{RAN} and \gls{UE}. In this work we focus on the 5G \gls{CN} and briefly describe the functionalities of the main \glspl{NF} that are crucial for the basic network operation.

\paragraph{Access and Mobility Function (AMF) and \gls{SEAF}} The \gls{AMF} is the UE entry point to the \gls{CN} for control messages. 
As such, \gls{AMF} is involved in UE registration, paging of idle UEs, and mobility management. Control messages 
are protected through encryption and integrity protection, and the \gls{AMF} stores the UE security context and decides which security algorithms the UE should use. Additionally, the \gls{SEAF} is commonly collocated with the \gls{AMF}, and acts as an intermediary between the \gls{UE} and the \gls{CN} in authentication procedures. 

\paragraph{Session Management Function (SMF) and \gls{UPF}}
5G follows a Control and User Plane Separation (CUPS) architecture, decoupling the UE control and data planes in two different \glspl{NF}: the \gls{SMF} and \gls{UPF} respectively. The \gls{SMF}
provides session management for the UEs, and
is responsible for establishing the PDU session between the UE and a given Data Network (DN), tunnelled through the \gls{UPF}. Hence, the \gls{UPF} is the interconnection point between UEs and other DNs outside of the cellular infrastructure and is in charge of encapsulation and decapsulation of UE traffic, packet routing and forwarding and QoS handling.

\paragraph{\gls{UDM} and Authentication Server Function (AUSF)}
The \gls{UDM} is a centralized database that stores UE information and credentials. The \gls{UDM} retrieves subscription data upon request, e.g. an SMF might request UE allowed services when configuring a PDU session. 
The \gls{AUSF} plays an important role in the UE authentication process. When a UE triggers a registration request, the \gls{AUSF} requests the authentication vectors from the \gls{UDM}, performs mutual authentication with the UE, and notifies the \gls{UDM} of the outcome. As a reference, \gls{AUSF} and \gls{UDM} functions together cover most of the functionalities of the Home Subscriber Server (HSS) of previous network generations.

\paragraph{\gls{NRF} and \gls{NEF}} The \gls{NRF} is the repository of all the \glspl{NF} and services currently available in the operator's network. As such, \gls{NRF} is in charge of: a) \gls{NF} registration, b) authorization and authentication  between NFs, and c) service discovery, i.e. exposing which services an \gls{NF} offers to the rest of the network. The \gls{NEF} exposes 5G \gls{CN} services and capabilities to third parties. \gls{NRF} and \gls{NEF} can be used together to provide a platform for 3rd party developers to create new services.

\paragraph{Network Slice Selection Function (NSSF) and Network Slice Specific Authentication and Authorization Function (NSSAAF)} Network slicing is a technique that enables the multiplexing of \glspl{NF} and network interfaces to create an isolated end-to-end network that fulfils certain requirements, e.g. URLLC and eMBB.
During initial access, the UE accessing the network requests a set of slices. This information is forwarded to the NSSF, which confirms the selected slices, or assigns a default slice if the UE did not specify any. The lack of slicing access control procedures in the initial Release 15 lead to unauthorized slice access by a UE. This was addressed by the inclusion of the NSSAAF in Release 16. The NSSAAF performs slice-specific authentication and authorization after initial registration to the CN.~\cite{TS_networkslicing_authorization}. 

For brevity, we direct the reader to the 3GPP documents~\cite{TS_rel17_7} for details about the remaining \glspl{NF}.

\subsection{Service-Based Architecture in the 5G Control Plane}
\begin{figure}
    \centering
    \includegraphics[scale=0.22]{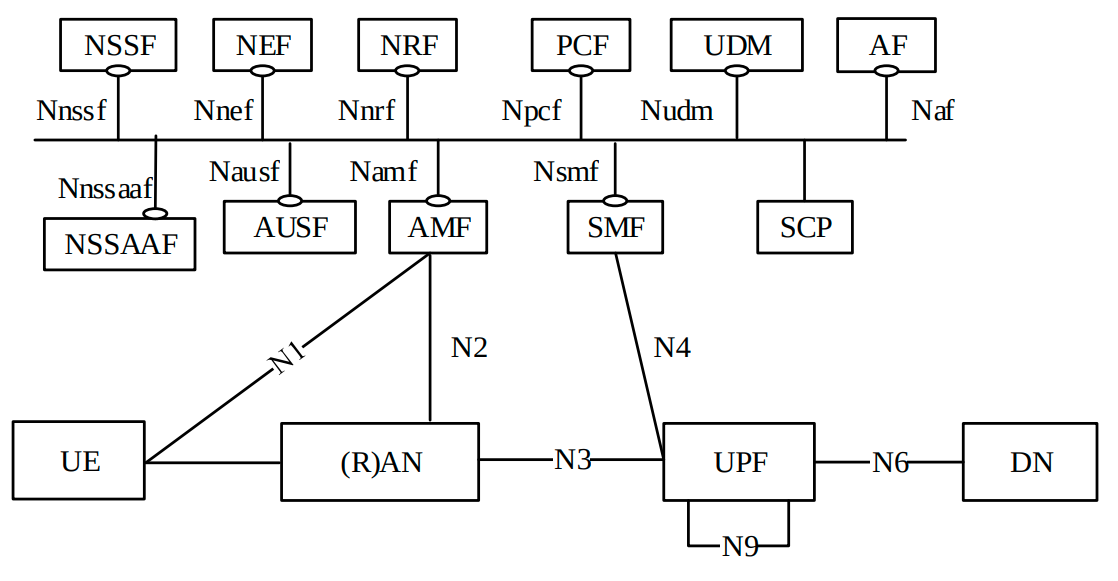}
    \caption{5G System Architecture~\cite{TS_rel17_7}}
    \label{fig:5g_system_architecture}
\end{figure}
In order to enable versatile and agile network configurations (e.g. relocate NFs to other servers) the control-plane \gls{NF} interfaces need to evolve. The traditional model of point-to-point connections between physical entities is inefficient for dynamically reconfigurable \glspl{VNF}. 5G adapts to the new paradigm with a Service-Based Architecture for the control-plane interfaces. 

This architecture follows an \gls{NF} Service Framework, where network functionalities are exposed as on-demand web services by an authorized NF (NF Service Producer) to an authorized \gls{NF} (NF Service Consumer) through a \gls{SBI}. The interfaces exposed by \glspl{NF} can be seen in Figure~\ref{fig:5g_system_architecture}, where for instance \textit{Namf} is the interface exposed by the \gls{AMF} to other service consumers. This way, different vendors can deploy \glspl{NF} that adhere to the \gls{SBA} without modifications to the existing system. Moreover, the 5GC control-plane uses modern software frameworks, i.e. RESTful design of APIs, JSON and HTTP/2 over TCP for the \gls{SBI}.

\subsection{Management, Orchestration, and Automation}
Management and Orchestration of a cellular network entails a diverse set of functions that include service provisioning, network configuration and planning, monitoring and fault management, billing and subscriptions, revenue management, product marketing, and more. Traditional communication networks managed these activities with two separate IT systems, \gls{OSS} and \gls{BSS}. OSS and BSS were distinguished by a separation of concerns between network operations and the business plan around which the network is built. With an increasing number of services provided by cellular networks, \gls{OSS}/\gls{BSS} are now commonly referred to as one system, and are connected by simple interfaces. To support the VNF infrastructure and automated 5G core deployment, ETSI introduced the \gls{NFV} \gls{MANO}~\cite{ETSI_NFV_MANO} layer, which was subsequently adopted by 3GPP~\cite{TR_NFV_MANO}.

OSS/BSS and NFV MANO are closely interconnected to support static allocation and dynamic reconfiguration of network functions and slices, based on the requirements of the network. Substantial work in industry and academia seeks the optimal management and orchestration model, and which layers should be responsible for static or dynamic allocation of \glspl{NF} and \glspl{NS}~\cite{mano_impacts_on_nfv, slice_management_MANO_OSS, flinck-slicing-management-draft, network_slicing_orchestration, onap_e2e_slicing, cisco_slicing, 3gpp_mano_charging, nesmo_paper_slicing}.
The OSS/BSS and MANO architecture details are beyond the scope of this work, however their role in automated deployment of the core is taken into consideration in our proposed model and security analysis.

\subsection{Beyond 5G (B5G) Networks}
The \gls{5GC} architecture was designed with future cellular networks in mind, commonly referred to as \gls{B5G} networks.
6G and 7G are predicted to benefit significantly from the efficiency, flexibility, and automated-reconfiguration potential, as most of the changes they introduce affect the \gls{RAN}.
In the rest of this section we provide some details on various improvements and changes expected in B5G networks.

The sixth generation of cellular networks is proposed to integrate advanced features in the existing
5G technologies.
6G will build on the 5G infrastructure in order to expand on land, air, sea, and space~\cite{isar_linkedin}. Some of the 6G milestones in
communications are holographic communications, artificial intelligence, high precision manufacturing, new technologies such as sub-THz or VLC (Visible Light Communications), 3D coverage framework, terrestrial and aerial radio Access Points. To this end, 6G is planned to transform the \gls{RAN} architecture into a cell-less model, where the UEs connect to the \gls{RAN} and not
to a single cell~\cite{5g_vs_6g}, and take advantage of higher frequencies than the ones used in 5G.
Moreover, 6G is planned to virtualize additional components of the PHY and MAC protocol layers.
Currently, PHY and MAC require dedicated hardware implementations as was the case
for the core network before 5G, but operators and vendors are already moving toward virtualized components~\cite{ibm_what_is_vran, samsung_vran}. 6G is expected to standardize this in a model architecture, similarly to the 5G \gls{CN}. This virtualization will decrease the costs of networking equipment and make massively dense deployment in 6G economically feasible. For the time being, 6G is merely a symbol, with a single definition or standard not existent, and an expected launch date by 2030~\cite{dficlub_5g_vs_6g}.

Subsequent technologies like 7G are expected to add even more features on top of existing versions, such as
satellite functionalities, intelligent radios, higher frequencies, and extensive use cases such as crime control, health monitoring, disaster preparedness, and more~\cite{future_mobile_technologies, 7g_comms}.
One important observation is that modern and future cellular networks follow an architectural
paradigm set by 5G, with most of the expected advancements happening in the \gls{RAN}, and the mobile
core being extended as needed. It is therefore even more critical to invest in future-proofing
the mobile core security in response to the very recent architectural paradigm shifts from 4G to 5G.

\section{Security in B5G Networks} \label{sec:sec_model}
In this section we provide an overview of the security mechanisms utilized at different layers of 5G networks. We analyze the security implications of the transition to a service-based, decentralized architecture, and describe our problem statement.

\subsection{Security overview in 5G Core}
\noindent The increasing importance of security in cellular networks is embodied by 3GPP in the new set of security requirements defined for 5G in Technical Specification 33.501~\cite{TS_sec_rel16}. The evolution of the 5G \gls{CN}, now fully virtualized and built on top of web services deployed on cloud-based architectures, requires a completely new security approach. Furthermore, the incorporation of an \gls{SBA} approach requires careful access control management, as now \gls{NF} communications are no longer point-to-point but service-based, i.e. available to everyone, and any \gls{NF} can request services on-demand. Hence, the 5GC follows a multi-level security approach for the control plane.

At the transport layer, 3GPP mandates that direct communication between NFs supports mutual authentication and data confidentiality and integrity, which is addressed by the use of TLS v1.2 or v1.3. In addition, access authorization for \gls{NF} services at the application layer is achieved using OAuth2.0, with the \gls{NRF} performing the role of the authenticator. TLS and OAuth2.0 are supported by a Public-Key Infrastructure (PKI) for the client and server certificates, which requires provisioning of keys, a Certification Authority (CA), and secure storage of certificates in \glspl{NF}.
This aspect is not defined by the 3GPP standard, and a solution is up to operators and manufacturers.

Security in 5G roaming scenarios is enhanced by the addition of an edge function between mobile networks, the Security Edge Protection Proxy (SEPP). The use of SEPP adds end-to-end application level security, which was not provided by the Diameter Edge Agent (DEA) deployed in 4G roaming scenarios. The interface between the SEPP of a Visited PLMN (VPLMN) and that of a Home PLMN (HPLMN), and uses HTTP2 protected by TLS in the control plane. The SEPP serves as a Proxy for the Network Functions between Service-Based Interfaces of separate PLMNs, enabling the SBA across PLMNs while preserving the desired security guarantees.

\subsection{5G Architecture Security Challenges}
The transition to a virtualized 5G architecture
increases the network's attack surface, mainly by shifting from dedicated,
isolated hardware to general-purpose servers and cloud deployment.
Compromised execution elements, such as the virtualization hypervisor and the host Operating System (OS), implicitly weaken the trust on the confidentiality and integrity of data, code, and the general-purpose software and hardware stacks running on the network device.
Private clouds, hosting other, untrusted services, may also be affected by security compromises, such as a bug in the hypervisor, and the implications for the host and guests are still being studied~\cite{empirical_virtualization_vulnerabilities, virtualization_challenges}. T-Mobile and Verizon expressed concerns in trusting public clouds, whereas AT\&T relies on Microsoft Azure Cloud~\cite{att_ms_azure, verizon_tmob_private_cloud}.
In the following, we identify some newly introduced threats to the 5G Core Network.

\paragraph{Shared memory} Securely storing sensitive data in the memory of \glspl{NF} during processing raises
increasing concerns, because the physical memory is shared with the
host and other guest services. For example, the host has direct access to the public-private key pairs, which are used in the TLS handshake, or the long-term key pairs which are stored in the \gls{UDM} virtual memory~\cite{TR_virtualization}.

\paragraph{Cloud-related security challenges} Cloud computing is possible because of virtualization in storage, networking, and computing. These software-based solutions allow for multiple \glspl{VM} to
reside on the same physical infrastructure. However, this raises security issues with the host OS, hypervisor, VM management, and virtual networking~\cite{cloud_virtualization_security_issues, cloud_virtualization_security_challenges}. Multiple cross-VM and memory deduplication attacks have been experimentally demonstrated when VMs share physical resources~\cite{cross_vm_attacks, get_off_my_cloud, memory_dedup_attacks}.

\paragraph{Homogeneous software solutions} The adverse effect of Zero-Day exploits is greatly amplified when NFs share the same building blocks and libraries (e.g. OpenStacks framework). As a result, a single zero-day exploit would affect large part of the \gls{NF} infrastructure~\cite{TR_virtualization}.

\paragraph{Compromised virtualization layer}
A compromised \gls{VM} hypervisor can inspect and edit the memory of guest \glspl{VM}, access secret keys, and modify
data or their functionality. It may tamper with the virtual clock in order to affect certain
cryptographic operations in the VMs, such as pseudo-randomness generation. Finally, a hypervisor breakout
attack can lead to the compromise of other \glspl{NF} running on the same virtualization layer~\cite{TR_virtualization}.
In~\cite{ormandy2007empirical}, the authors perform an empirical study and discover multiple vulnerabilities in the
presence of a compromised virtualization layer against all tested VMs.

\paragraph{\gls{MANO} framework}
The \gls{NFV} \gls{MANO} combined with \gls{OSS}/\gls{BSS} control all \glspl{NF} deployment and life cycle management.
A compromised \gls{MANO} effectively compromises all network deployment: \glspl{NF} can be arbitrarily instantiated,
terminated, or migrated. Additionally, depending on the deployment strategy, \gls{MANO} may hold \gls{NF} credentials and
access privileges in order to deploy, manage, and migrate the \glspl{NF}. This could lead to unauthorized lateral movements
by compromising critical entities, such as the \gls{NRF}, and to unauthorized access of \gls{NF} data.

\paragraph{Secure storage of sensitive data} The above examples focus on the security of data while
in-use by the NFs, and temporarily exposed in the host memory. Similar issues arise when sensitive
data need to be stored in remote storage for longer periods of time. This includes customer or network
related data, credentials, temporary and long-term secret keys, certificates, and more.

\paragraph{Secure handling of permanent 5G identifiers} In an
attempt to reduce the exposure of permanent device identifiers and mitigate
tracking attacks, 5G introduces concealed versions of \gls{RAN} permanent identifiers (e.g. \gls{SUCI} replacing
\gls{SUPI}). However, after being de-concealed in the 5G \gls{CN},
the permanent user identifiers are stored in the \gls{NF} memory.
This is very reminiscent of a perimeter-based approach and suffers from several
disadvantages presented in section~\ref{subsec:per_based}.

\subsection{Problem Statement}
The above security analysis demonstrates the need to define a specific adversarial model.
\paragraph{Adversarial Model}
In the context of 5G \gls{SBA} and \gls{B5G} networks, where \glspl{NF} are virtually
deployed in remote servers or cloud infrastructure, we consider the case that the
infrastructure is untrusted, or the operator is not fully trusted (e.g. during roaming).
An adversary may be able to achieve a combination of:

\begin{itemize}
    \item compromise the virtualization layer of one or more NFs (e.g., using zero-day exploits)
    \item access and modify virtual resources
    \item gain access to the hardware infrastructure
\end{itemize}

To compensate for the fact that physical, specialized hardware, is replaced by off-the-shelf, shared
resources on the cloud, we must consider the execution environment untrusted, which includes the
remote hardware and software stack, excluding the \gls{NF} code.

\paragraph{Problem Statement}

Current 5G security mechanisms achieve secure communications between mutually trusted
network components, but are not sufficient to create an \textit{attestable} environment,
trusted to securely handle the data during execution.

\textit{There is a need for new mechanisms to ensure secure computation by mutually untrusted components of B5G networks,
built on limited root-of-trust points.}

\section{Zero Trust Architecture}
\label{sec:zta}

In the following, we discuss the relevance of a \gls{ZTA} design in the context of modern cellular networks. We point out the existing support for \gls{ZTA} principles in the 5G standard, and motivate the need to tie these principles with the underlying execution environment. To that end, we introduce \textit{Zero Trust Execution}, a set of \gls{ZT} tenets to model computation on untrusted execution environments.

\subsection{From perimeter-based security to the Zero Trust model}
\label{subsec:per_based}
Traditional systems relied upon perimeter-based security: a strong perimeter around the on-premises network
is established using network firewalls or Virtual Private Networks, and activities within this perimeter
are considered ``safe''. Modern systems include service endpoints in public or private clouds, as well as partners'
private infrastructure, forming a \textit{hybrid cloud architecture}~\cite{ibm_hybrid_cloud}. The security perimeter expands
to include not only the organization's local networks, but every access point that hosts, stores, or accesses corporate assets.
Maintaining a strong network perimeter that includes these diverse computation environments proves to be complicated,
expensive, and vulnerable; overall insufficient.~\cite{ms_perimeter_obsolete, ibm_zta} There is a need for a security model
that does not assume a threat-free environment, and handles the complexity of mobile, cloud, or distributed workloads, with strict verification policies and automated security controls.
A recently revived security approach
that aims to mitigate such threat models is \textit{Zero Trust (ZT)}~\cite{marsh_formalizing_trust,forrester_zero_trust_2010}.
The \gls{ZT} model replaces perimeter-based, one-time authentication,
with a ``\textit{never trust, always verify}'' mindset to tackle this challenge.

However, it is hard to derive a universally accepted security definition of Zero Trust; different use cases such as
cloud computing~\cite{ibm_zt_ciso_perspective} or international banking~\cite{ibm_zt_cib_study} have different
security and business requirements, and adopt ZT in different ways. Zero Trust is not a tool or an action, rather a
concept~\cite{wired_what_is_zt}, and its adoption is a phased, step-by-step process, that depends on the goals and requirements
of each and every organization, and its \gls{ZT} maturity~\cite{ms_zt_maturity}.
\gls{ZT} approaches generally suggest that certain assets must be secure in a setting where every connection, and every
endpoint are considered a threat. This is particularly challenging in cloud and multi-vendor situations~\cite{network_architectures} where
new entities try to access corporate assets from unrecognized locations, and \textit{context} is essential in order to classify the action as
suspicious or safe~\cite{zt_requires_context}.
In the recent years, security organizations are trying to identify the requirements of \gls{ZT}, \gls{ZTA}, and related principles~\cite{network_architectures, nist_zt_doc, nist_zta}, with
proposed solutions in network, data, and cloud security~\cite{ibm_zta}.

\subsection{Zero Trust in B5G Networks}
\begin{figure}
    \centering
    \includegraphics[width=0.9\linewidth]{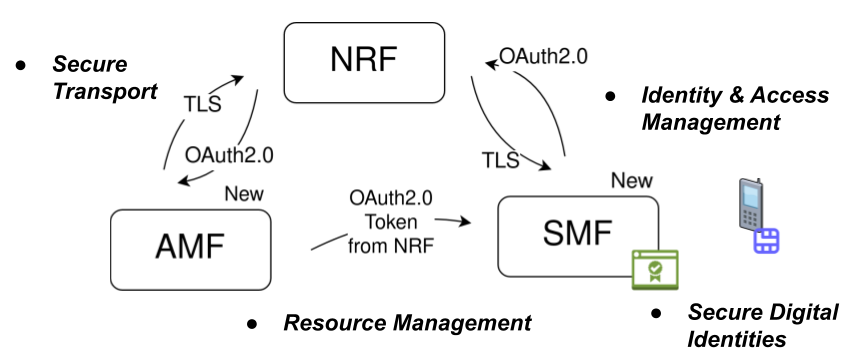}
    \caption{Zero Trust principles in 5G networks: secure digital identities, secure transport, resource management}
    \label{fig:5g_principles}
\end{figure}

Manufacturers like Ericsson have already reviewed the support for \gls{ZT} in the 5G standard, in network access security,
network domain security, and SBA domain security~\cite{ericsson_5g_zero_trust}.
The four key \gls{ZT}-enabling security features are secure digital identities, secure transport, policy frameworks,
and secure monitoring. Some examples are shown in Figure~\ref{fig:5g_principles}, and
include SIM cards and X.509 certificates for secure digital identities and
access management, TLS for secure transport, and OAuth2 tokens for resource
management. The 5G standard additionally suggests and specifies constant monitoring and policy frameworks enforcement.

However, \gls{ZT} principles alone are not sufficient to reinforce trust in the network, because they crucially depend on the execution environment of the deployment. In previous generations, the use of dedicated, audited software and hardware components, and the monitoring of physical access to resources, attested to a trusted execution environment.
In contrast, in the new 5G architecture, the underlying execution environment is considered part of the threat model, which jeopardizes the correct evaluation of 5G \gls{ZT} principles.
For instance, the \gls{ZT} principle of secure identities (e.g. digital certificates) offers lower trust value, because the augmented threat model makes it more vulnerable to compromise (e.g. direct memory access by a malicious hypervisor).

To address this, we aim to restore trust to the execution environment, and formalize the principles
that can attest to it. 
Confidential computing technologies will be critical in this effort~\cite{ericsson_5g_zero_trust, ericsson_paper_building_trust}, with hardware-based roots of trust forming the basis for security
assurance in our work. We discuss confidential computing solutions in more detail in a following section.

\subsection{Vertical Zero Trust} \label{subsec:vertical_zt}
In this work, we extrapolate the basic tenets of \gls{ZT} ``vertically'', from the network enterprise layer to the
lower software and hardware layers of the execution stack.
By following those tenets we strengthen the trust in the execution environment, achieving \textit{\gls{ZTE}}.
The tenets are designed from the perspective of what should be followed, rather than what should be avoided.
Even though adhering to all \gls{ZTE} tenets is the ideal goal, any subset of them can be independently adopted
alongside perimeter-based defenses in hybrid models, depending on the deployment needs.
Further, they are technology-agnostic, and can be realized using various potentially orthogonal techniques.
The following tenets are a natural extension of the ZT tenets
defined by NIST in~\cite{nist_zt_doc}, and both sets can complement one-another in order
to achieve a robust ZTA.

\textit{A Zero Trust Execution (ZTE) is deployed with adherence to the following zero trust
basic tenets:}

\begin{itemize}
    \item \textbf{All code, data, and computation sources in the execution are considered (execution) resources.} The execution may involve
    parts of code from different vendors, specialized or off-the-shelf hardware components, and enterprise, client, or third-party data.
    The execution deployment may comprise private or enterprise-owned infrastructure, public or private cloud platforms, or a hybrid of the above.
    \item \textbf{All executions must be secured regardless of deployment platform: on-premises network alone does not imply trust.} 
    Execution performed on privately owned infrastructure must meet the same security requirements as executions deployed on public cloud or similar.
    Trust should not be automatically granted based on the execution being deployed on enterprise components. The complete execution stack must
    be protected in the most secure manner available, protect data and code confidentiality and integrity, and provide data and code source authentication.
    \item \textbf{Execution on individual enterprise execution resources is granted on a per-session basis.} Trust in the execution resources
    is evaluated before the access is granted. Execution should proceed using the least amount of computational (software and hardware modules) and
    informational (personalized or control data) resources required to complete the task. This can mean only "sometime recently" instead of directly
    before execution for this particular resource. However, access to one resource does not automatically grant access to a different resource.
    \item \textbf{Access to execution resources is determined by dynamic policy — including the observable state of code identity, application/service,
    and the execution task — and may include other behavioral and environmental attributes.} An organization protects resources by defining what resources
    it has, who its members are, and what access to resources those members need. For zero trust execution resources, members can be entities that need
    access to execution resources to complete a certain functionality. Policy is the set of access rules based on attributes that an organization assigns
    to a subject, data asset, or application. Policy related attributes can include member characteristics such as software versions installed,
    execution platform, previously observed executions, code source credentials and remote attestation. Behavioral attributes
    include automated analytics, execution history logs, measured deviations from observed execution patterns, and more. These rules and attributes
    are based on the needs of the business process and acceptable level of risk. Resource access and action permission policies can vary based on the
    sensitivity of the resource/data. Least privilege principles are applied to restrict both visibility and accessibility.
    \item \textbf{The enterprise monitors and attests the security posture of all owned and associated assets.} 
    No asset is inherently trusted. The enterprise evaluates the
    security posture of the asset when evaluating an execution request. An enterprise
    implementing a ZTE should establish a continuous diagnostics and mitigation (CDM) or
    similar system to monitor the state of execution resources and should apply
    patches/fixes as needed. Assets that are discovered to be subverted, have known
    vulnerabilities, and/or are not managed by the enterprise may be treated differently
    (including denial of all connections to enterprise resources) than resources owned by or
    associated with the enterprise that are deemed to be in their most secure state. This may
    also apply to associated resources (e.g., remote cloud instances) that may be allowed to
    access some resources but not others. This, too, requires a robust monitoring and
    reporting system in place to provide actionable data about the current state of enterprise
    resources.
    \item \textbf{All resource authentication and integrity are dynamic and strictly enforced before execution is performed.}
    This is a constant cycle of obtaining access, scanning and
    assessing threats, adapting, and continually reevaluating trust in ongoing execution.
    An enterprise implementing a ZTE would be expected to have code integrity, data
    confidentiality and integrity, and resource management systems in place.
    Continual monitoring with possible reevaluation occurs
    throughout execution, as defined and enforced by policy (e.g., time-based, new
    resource requested, resource modification, anomalous subject activity detected) that
    strives to achieve a balance of security, availability, usability, and cost-efficiency.

    \item \textbf{The enterprise collects as much information as possible about the current state of assets, and deployment infrastructure and uses it to improve its security posture.} 
    An enterprise should collect data about execution security posture, execution requests, and data access,
    process that data, and use any insight gained to improve policy creation and enforcement. This data can also be used to
    provide context for execution requests from subjects. Attesting the environment facilitates these capabilities. 
\end{itemize}

An enterprise implementing ZTE should follow the above tenets and a set of assumptions, that depend on the design of the execution deployment, and the assets involved.
We present an example set of assumptions that, along with the \gls{ZTE} tenets, provides a view of a \gls{ZTE} in the context of the \gls{5GC}:
\begin{itemize}
    \item \textbf{The entire enterprise-owned deployment stack is not considered an implicit trust zone.}
    Assets should always act as if any component of the execution stack is compromised, and execution
    should be performed in the most secure manner available. This entails actions such as enforcing
    data confidentiality and integrity in-use, in-flight, and at-rest.
    \item \textbf{Parts of the execution stack may not be owned or configurable by the enterprise.}
    Execution components may be COTS hardware or third party libraries and OS may be configured by a cloud provider.
    B5G NFs may be developed and maintained by different vendors and application auditing services may be outsourced
    to third parties.
    \item \textbf{No resource is inherently trusted: Every asset in the execution is evaluated before participating in the execution.}
    Every asset must have its security posture evaluated via a Policy Enforcement Point before a request is granted to
    an enterprise-owned resource (similar to tenet 6 above for assets as well as subjects). This evaluation should be
    continual for as long as the session lasts. Resources may have artifacts that provide a confidence level higher than
    the same request coming from other resources. This can include security evaluation certificates, confidential computing
    attestations, and more.
    \item \textbf{Not all resources are deployed on enterprise-owned infrastructure.}
    Execution may be performed on remote enterprise deployments as well as cloud services. Enterprise assets
    may need to utilize the local (i.e., non-enterprise) components to complete a functionality such as
    computation resources, storage devices, or network interfaces. A B5G related functionality may need
    to utilize a local or foreign operator or cloud resources, for example virtual operators using another
    operator's physical infrastructure, or during roaming.
    \item \textbf{Remote enterprise assets cannot fully trust their deployment stack.}
    Remote assets should assume that the local (i.e., non-enterprise-owned) execution stack is hostile.
    Assets should assume that all execution resources can be accessed by the adversary. All processing
    requests should be authenticated and authorized, and all executions should be done in the most
    secure manner possible (i.e., provide data and code security, integrity protection, and source authentication).
    \item \textbf{Assets and workflows between enterprise and non-enterprise infrastructure should have a consistent
    security policy and posture.} NFs are
    highly interconnected to achieve core functionality, for instance AMF requesting data from UDM and forwarding
    data to SMF. These NFs may not all be deployed in similar infrastructure; assets and workloads should
    retain their security posture when moving from one type of infrastructure to another, like in the roaming deployment. This security model can be applied to handover as well, even though the critical assets are moved within the MNO's infrastructure. This includes workloads
    migrating from on-premises data centers to non-enterprise cloud instances.
\end{itemize}

\section{Trusted Execution Environments}
\label{sec:tees}
\noindent Confidential computing relies on hardware-rooted security and attestation to
produce an isolated execution environment, known as \gls{TEE}. Major chip manufacturers like Intel, AMD, ARM, or IBM,
deploy their own \gls{TEE} solutions~\cite{intel_sgx, intel_tdx, amd_sev, arm_trustzone_cortex_m55_paper, ibm_pef_paper}, and have proposed extensions that build secure computation platforms around their \glspl{TEE}.
Based on our adversarial model we consider only hardware-based \glspl{TEE}, that is, \glspl{TEE} that are built using hardware modules called \textit{enclaves}, rather than software-emulated solutions. An enclave is a physically isolated piece of hardware, usually part of a larger chip like a CPU, that is designed to perform specific operations.
Enclaves usually have their own instruction set architecture, access to specified parts of the memory,
and the ability to execute application code. During the past years, enclaves have
been an important primitive for hardware-based isolated execution environments for computation on
critical data, with applications in banking mobile apps, biometric authentication, protecting secret keys,
Point-of-Sale (POS), and Digital Rights Management (DRM).

The \gls{TEE} threat model considers as trusted only the application code, the enclave, and its manufacturer, which eliminates any trust placed in the intermediate layers during computation. As a consequence, the computation is considered secure in the presence of an untrusted hypervisor, operating system, virtualization layer, kernel, and hardware other than the enclave. Specifically, \glspl{TEE} guarantee the security of code and data in-use from the moment they are loaded in the enclave-protected system memory, while processed by the enclave, and up to their release.
They provide a secure input/output (I/O) channel between the enclave and the rest of the
system, and perform the critical parts of the computation inside the physically isolated enclave, only exposing
the final output to the rest of the system.
Other techniques can be used to protect data in-rest or in-flight, such as \gls{FDE} or transport layer security.
Nevertheless, \glspl{TEE} have been used as a means to strengthen the security of data in-rest by exposing the \gls{FDE} key only inside the enclave~\cite{sgx_storage_encryption, tresor_sgx_git}.
Finally, while some \glspl{TEE} present an extra layer of security against side-channel attacks~\cite{amd_sev_snp_paper},
certain cases such as architectural side channel attacks are not currently included in their adversarial model, and good-practice programming techniques are advised~\cite{amd_sev_snp_paper, intel_sgx_side_channel} to counter these attacks.

A vital property of any \gls{TEE} is \textit{attestation}, i.e. the ability of the \gls{TEE} to provide \textit{proof} of secure computation, upon request from a \textit{requesting entity}. Attestation returns a report of the complete system description, such as hardware configuration and software versions. The requesting entity verifies the system status before loading any sensitive information into the enclave and initiating secure computation.
\glspl{TEE} usually achieve that by an attestation mechanism that sends an intermediate report to the chip manufacturer infrastructure that plays the part of the \textit{verifying entity}. The report is verified, signed, and sent back to the enclave, that forwards it to the requesting entity. Usually, the report is accompanied by a secret key for a secure channel between the user and the enclave. The attestation can be either local, if the requesting entity needs to be located in the same server as the enclave, or remote, if there is no such restriction.

An in-depth comparison between available \glspl{TEE} is beyond the scope of this work, and the latest state of
affairs is constantly evolving due to technological advancements and company marketing strategies. We direct the
reader to the literature for a history of \glspl{TEE}, comparisons, and benchmarks~\cite{intel_tdx_paper, amd_sev_snp_paper, ibm_pef_paper, sgx_sev_comparison}.
However, we examine the two main categories of \glspl{TEE} in order to determine which is more suited to the use-case of
the 5G \gls{CN}. Depending on the size of the isolated computation environment they provide, called \textit{\gls{TCB}}, \glspl{TEE} can be classified into \textit{application-based} \label{app_based_tee} \glspl{TEE} and \textit{VM-based} \glspl{TEE}. The former provide a small \gls{TCB} consisting of a few hundred lines of code and libraries, that result in a smaller attack surface for the execution. This is tailored for use in embedded devices where a small part of an application requires very strict security guarantees, such as protection of long-term secret keys, credentials, or biometric data. The latter, provide a truly isolated \gls{SVM} with its own guest OS and libraries, that can accommodate larger applications. This creates a large TCB that includes tens of thousands of lines of code for the guest OS, packages, libraries, and applications, that increases the attack surface.

However, VM-based \glspl{TEE} have significant deployment advantages over application-based \glspl{TEE}, making them preferable for deployment in cloud computing, and large-scale and decentralized applications running on data centers. For instance, they provide the ability to encapsulate an entire large application in an \gls{SVM} with a simple BIOS configuration and no application code refactoring. On the contrary, application-based \glspl{TEE} require major refactoring in the application code, which makes it highly non-trivial to achieve, especially for legacy applications~\cite{sgx_sev_comparison}.
Further, because of their limited physical memory access, protected memory must be released before loading subsequent parts of the code inside the enclave. Finally, some application-based \glspl{TEE} such as Intel SGX, prevent the enclave from executing system calls, and interrupt the enclave execution flow every time a system call is required. Even though there are frameworks that facilitate the translation from application code to enclave instructions, and amortize the system calls overhead~\cite{scone_paper}, the impact on performance is still significant. This makes application-based \glspl{TEE} overall inefficient for larger applications with many system calls.

\begin{figure}[t]
    \centering
    \includegraphics[width=0.9\linewidth]{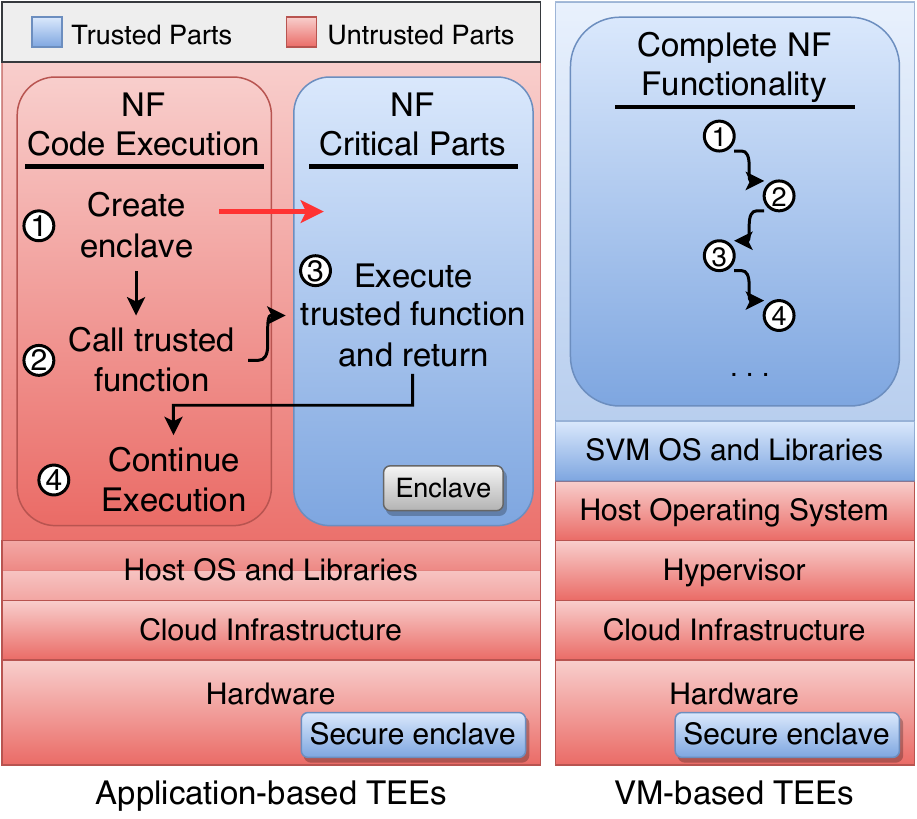}
    \caption{High level NF functionality in application-based vs VM-based \glspl{TEE}}
    \label{fig:enclave_types}
\end{figure}

Figure~\ref{fig:enclave_types} shows a side by side comparison of a 5G \gls{NF} deployment using application-based TEEs, and VM-based TEEs. The trusted and untrusted parts are highlighted in blue and red respectively. In the case of application-based TEEs the NF application code is split in two parts. The security-sensitive part is executed in the enclave, and the execution flow moves between the CPU and the enclave whenever necessary. In the case of VM-based TEEs, the entire NF application is encapsulated inside an \gls{SVM}, with no changes to the code or execution flow. The SVM may be managed by a Type I (bare-metal) hypervisor or a Type II (hosted) hypervisor on a host OS. In both cases, the hypervisor and the optional host OS are considered untrusted.

In the context of distributed, virtualized \glspl{NF}, application-based \glspl{TEE} would require major code refactoring, would incur serious performance overhead, and would not \textit{fit} the scale of the functionality inside small enclaves.
VM-based \glspl{TEE} are better suited for this purpose; we expand on that in Section~\ref{sec:results}, where we evaluate how the above solutions address different use case scenarios within the new 5G architecture.

\subsection{AMD Infinity Guard} 

Because of its maturity and wide deployment, the AMD \gls{SEV}~\cite{amd_sev} of the AMD Infinity Guard family of products, is of particular interest.
AMD has partnered with multiple cloud providers~\cite{amd_sev_cloud_partners}, and is the first VM-based \gls{TEE} to provide many desired features such as data and code confidentiality and integrity, and remote attestation.

The basic building block for AMD \gls{SEV} is \gls{SME}~\cite{amd_memory_encryption_paper}, a mechanism that encrypts parts of the system memory, creating the AMD enclave, called \gls{SP}.
AMD \gls{SEV} encrypts the whole memory image of every \gls{SVM}, and supports multiple \glspl{SVM} with different encryption keys each. Recent \gls{SEV} extensions, \gls{SEV}-\gls{ES}~\cite{amd_sev_es_paper} and \gls{SEV}-\gls{SNP}~\cite{amd_sev_snp_paper} provide additional security guarantees such as encrypted CPU registers, better protection against side-channel attacks, improved remote attestation, and crucially, data integrity for the \glspl{SVM}. AMD Infinity Guard tools are deployed in the AMD EPYC line of processors for data center use, with the EPYC 3rd generation supporting up to the latest \gls{SEV} extension.

AMD has published general-purpose benchmarks for AMD SEV~\cite{azure_sev_bench} that demonstrate very low performance overhead over the execution outside the TEE. AMD attribute this to an extremely fast, hardware-accelerated, 128-bit AES encryption module that is embedded in the memory controller and is used for memory encryption by AMD SEV.

\subsection{Alternative TEE solutions}

Alternatives to AMD \gls{SEV} are not mature enough to be included in our benchmarks. However, as these platforms evolve, they may prove viable for deployment in the context of \gls{B5G} networks in the near-term future.

\paragraph{Intel Platforms}
Intel was the first manufacturer to introduce secure enclaves in
their processors with the Intel \gls{SGX}~\cite{intel_sgx} in 2015. \gls{SGX} is an
application-based \gls{TEE} that supports multiple enclaves at the same time with up to 128 MB
total physical enclave memory, a limitation lifted in \gls{SGX}2.0 at the expense of data integrity.
Intel \gls{SGX} was recently deprecated from consumer CPUs 11th generation or newer, but is still included in
data center solutions such as the Intel Xeon processors. Intel proposed the \gls{TDX} in 2019~\cite{intel_tdx, intel_tdx_paper},
a VM-based \gls{TEE} that uses \gls{SGX} to achieve \gls{TME}, similar to AMD's VM-based solutions.
This move continues the trend toward VM-based \glspl{TEE} that offer flexible deployment, and use enclaves as their basic primitives. Intel TDX is planned for deployment on Intel Xeon 4th generation processors.

\paragraph{ARM TrustZone and \gls{CCA}} ARM offers their own secure computation solution in their Cortex-based
processors~\cite{arm_trustzone_cortex_a, arm_trustzone_cortex_m}, an application-based \gls{TEE} that creates two separate environments that run on the same
CPU core. The \textit{trusted world} and the \textit{untrusted world} are connected with a secure I/O channel, i.e. a TrustZone software instruction, the Secure Monitor Call (SMC).
Essentially, ARM TrustZone provides two virtual processors with their own dedicated memory, backed by hardware-based
access control. Applications can then switch to the trusted world to perform small, security sensitive
tasks, and use the untrusted world for the main functionality.
TrustZone is deployed in many processors~\cite{arm_trustzone_cortex_m55_paper, arm_trustzone_fido_paper}, but their proprietary implementation and non-disclosed implementation details complicate their security evaluation.

ARM expanded to the VM-based TEE domain with \gls{CCA}~\cite{arm_cca} in ARMv9, a set of hardware and software solutions that enable the creation of dynamic and attestable \glspl{TEE}, called \textit{Realms}. ARM \gls{CCA} is an isolation technology that builds on the existing TrustZone technology, but is optimized for large, compute-intensive workloads, expected to reach production in 2024~\cite{arm_cca_fuse_wiki}.

\paragraph{IBM \gls{PEF}}
IBM \gls{PEF}~\cite{ibm_pef_paper} recently joined the VM-based confidential computing solutions in 2021, supporting multiple \glspl{SVM}, similarly to the AMD and Intel VM-based \glspl{TEE}.
IBM \gls{PEF} partitions the \gls{TEE} between hardware and firmware in a novel way: it achieves SVM isolation by hardware-based access control only, and uses existing cryptographic tools like the Trusted Platform Module (TPM) for data confidentiality and integrity. Further, \gls{PEF} omits remote attestation to simplify the SVM life cycle management, and leverages local attestation to achieve security.
This is achieved via a new entity, the Protected Execution Ultravisor that has higher privileges than the hypervisor, and is responsible for managing the hardware-based cryptographic operations and access-control. If a step of the VM verification fails during initiation, local attestation fails and the SVM is rejected.

The lack of memory encryption in IBM PEF leaves it vulnerable against attacks like boot-time memory probing, but leads to minimal impact on computation. However, \glspl{SVM} in PEF are heavily impacted by small message exchanges in terms of throughput, leading to nearly 45\% degradation~\cite{ibm_pef_review}. This is probably due to the overhead in the secure I/O path, and the context switching between the SVM and the host. Even though this degradation drops to 10\% as the message size grows, IBM PEF might not be an ideal solution for the small-size control messages exchanged in the 5G CN until this limitation is lifted.
IBM PEF is deployed in the latest generation of POWER9 processors, but is scarcely available by cloud providers.

\subsection{NF-to-TEE Migration Challenges}
We acknowledge that AMD SEV is only supported by specific AMD processors, and a complete migration to different infrastructure might be costly for a network operator. In this section, we determine the viability of securing \glspl{NF} using \gls{SGX} enclaves. To achieves this, our threat model requires the complete implementation of an \gls{NF} to be executed inside an enclave. This way, existing Intel-based infrastructure can be leveraged in our system design alongside AMD \gls{SEV}. We analyze the functionalities of 5G \glspl{NF} as described in the standard, as well as current open-source implementations, to evaluate their candidacy for efficient deployment on Intel \gls{SGX}. We define the following metrics that describe various SGX limitations:
\begin{itemize}
    \item Metric 1 (M1): The NF does not require constant transfer of high amounts of data.
    \item Metric 2 (M2): The NF does not manage a large database.
    \item Metric 3 (M3): The NF does not perform latency-critical operations.
    \item Metric 4 (M4): The NF does not require complex code refactoring.
\end{itemize}

\begin{table}[t]
\caption{Viability analysis of deploying NFs on Intel SGX.}
\label{tab:tableSGX}
\begin{tabular}{l|cccccccc|}
\cline{2-9}
                         & \multicolumn{1}{l|}{AMF} & \multicolumn{1}{l|}{SMF} &  \multicolumn{1}{l|}{AUSF} & \multicolumn{1}{l|}{UDM} & \multicolumn{1}{l|}{NRF} & \multicolumn{1}{l|}{UPF} & \multicolumn{1}{l|}{PCF} & \multicolumn{1}{l|}{NSSF} \\ \hline
\multicolumn{1}{|l|}{M1} &          \checkmark                &       \checkmark                   &    \checkmark                      &         \checkmark                  &             \checkmark             &          X                &           \checkmark               &    \checkmark                       \\ \cline{1-1}
\multicolumn{1}{|l|}{M2} &      \checkmark                   &        \checkmark                  &         \checkmark                 &            X               &     \checkmark                     &      \checkmark                    &           X               &  \checkmark                         \\ \cline{1-1}
\multicolumn{1}{|l|}{M3} &          X                &          \checkmark                &           \checkmark               &            \checkmark               &        \checkmark                  &         \checkmark                 &      \checkmark                    &  \checkmark                         \\ \cline{1-1}
\multicolumn{1}{|l|}{M4} &          X                &          X & \checkmark    & \checkmark & \checkmark & \checkmark &     \checkmark                     &  \checkmark                        \\ \cline{1-9}
\multicolumn{1}{|l|}{Res.} &          X                &          X & \checkmark    & X & \checkmark & X &     X                     &  \checkmark                        \\ \cline{1-9}
\end{tabular}
\end{table}

Table~\ref{tab:tableSGX} contains a summary of our analysis for the most relevant 5G NFs that we introduced in Section~\ref{sec:5gnfs}.
For M1, we determine that signalling and control messages do not constitute a bottleneck because they generate a low volume of data. On the other hand, user-plane functions need to process all the data coming from/to the UE and analyze QoS metrics. These tasks require a considerable amount of data to be processed inside the enclave.
For M2, we observe that certain \glspl{NF} manage databases that scale in size with the number of users in the network, and deploying these \glspl{NF} on \gls{SGX} would be inefficient and cumbersome. For instance, UDM contains important information such as user credentials, and PCF manages the charging policies for each \gls{UE}. Databases that do not scale with the number of users are more suitable candidates for SGX deployment. For instance, the NRF contains NF information and keeps track of registered services and NFs, which scales with the number of NFs and services.
Regarding M3 and M4, we analyze the number and complexity of functionalities offered by each NF, from three perspectives: a) as defined in the 3GPP standard, b) as defined in the documentation of industry core network solutions~\cite{oracle5Gcloud}, and c) as coded in open-source \gls{5GC} implementations~\cite{Open5GS}.
We observe a large disparity in the code complexity and length of various \glspl{NF}, which is aligned with the functionality each one performs.
For instance, the AMF and SMF implementations span more than 20000 lines of code each, and involve communication with the UE and gNB, i.e. NAS and NGAP protocol management respectively. This makes code refactoring complicated for these NFs.
Other NFs have less than 5000 lines each, such as the NSSF, which is used to retrieve and update slice information, a limited and fairly simple task.
Finally, the AMF performs latency-critical procedures like UE authentication or handover handled by the AMF, which would be affected by the SGX overhead.

\section{System Design and Analysis}\label{sec:system_design}
\noindent In this section we discuss our proposed model architecture for the 5G \gls{CN} using AMD \gls{SEV} as a \gls{TEE}. The same design principles apply to future iterations of \gls{B5G} networks, which adopt a similar SBA and security model. They also naturally adapt to virtualized \gls{RAN} architectures, which are expected in \gls{B5G} networks and will likely follow a similar security model. We note that different \gls{TEE} solutions can be used with minimal changes to our design.

\subsection{Model Architecture}

\begin{figure*}[t]
    \centering
    \includegraphics[width=0.75\linewidth]{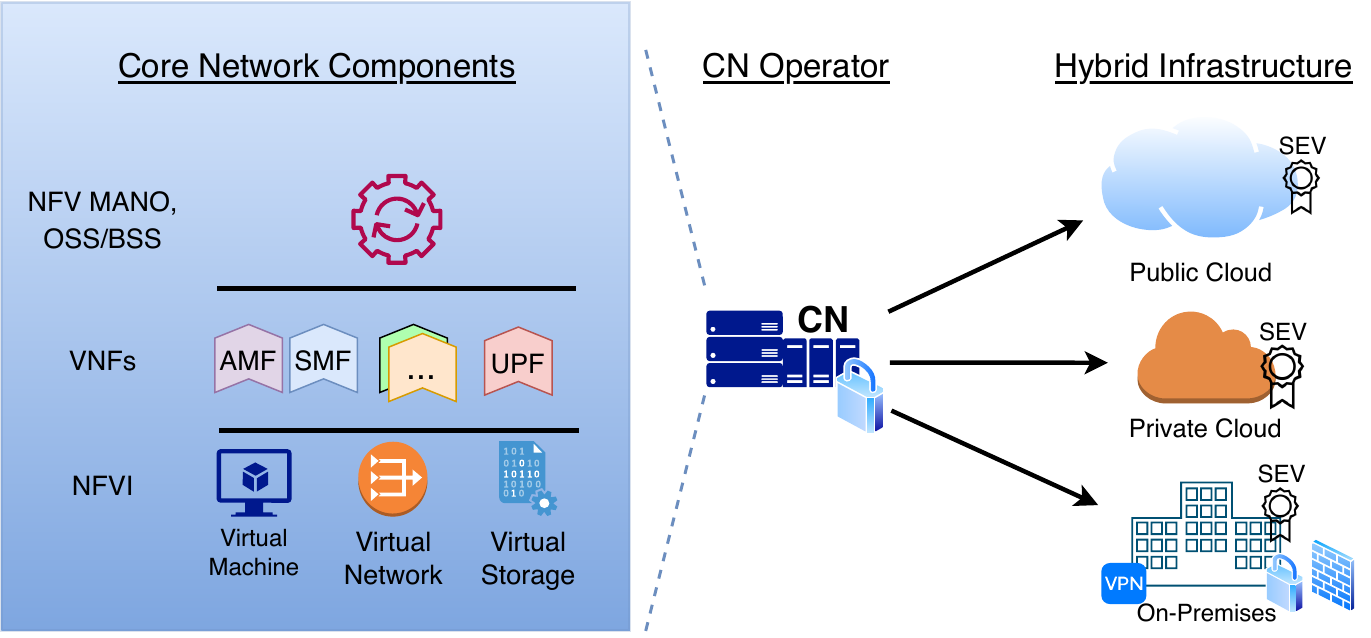}
    \caption{Confidential Core Network proposed architecture}
    \label{fig:proposed_model}
\end{figure*}

Our 5GC model architecture treats the COTS hardware and software, cloud provider, and hypervisor,
as untrusted. Their identities and validity are backed by digital identities and security
practices such as penetration testing, auditing, and formal methods of verification, but they are
considered prone to arbitrary security vulnerabilities at any point in the deployment timeline. To
enhance 5GC security we propose the following deployment architecture with no changes to the 5G standard by 3GPP, which is displayed in Figure~\ref{fig:proposed_model}.
\begin{itemize}
    \item The 5GC operator deploys \gls{CN} components on infrastructure that may not be operator-owned,
    on-premises, private cloud, public cloud, or a hybrid of the above. We refer to this platform as \textit{host infrastructure}.
    \item The \gls{CN} components comprise the \glspl{VNF}, the
    \gls{NFVI} and the \gls{NFV} \gls{MANO} and \gls{OSS}/\gls{BSS} layers.
    \item The host infrastructure supports AMD SEV-SNP, i.e. a bare-metal environment equipped with AMD EPYC 7xx3
    (Milan) processor.
    \item The \gls{CN} components adhere to the paradigms of \gls{NFV} and \gls{SBA} as introduced by the 5G standard.
    They may be developed and maintained by the 5GC operator or a third-party vendor.
\end{itemize}

\begin{figure*}[ht]
    \centering
    \includegraphics[width=\linewidth]{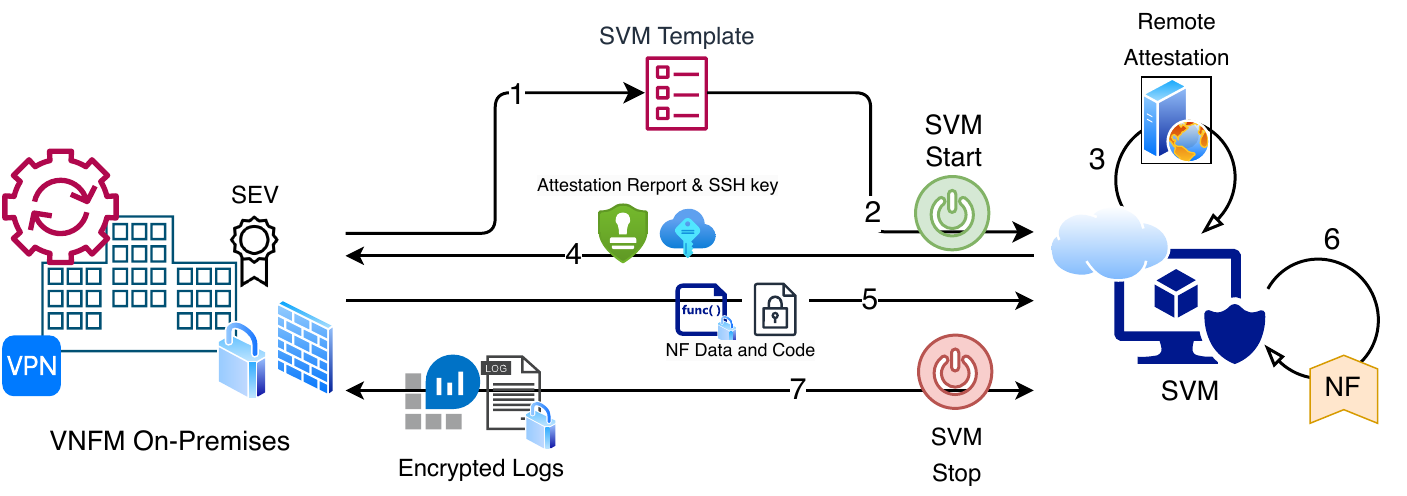}
    \caption{NF deployment and life cycle flow}
    \label{fig:deployment_flow}
\end{figure*}

The life cycle of an \gls{NF} in our architecture follows the 3GPP standard, with few modifications to take advantage of AMD SEV. We present a flow of a \gls{NF} deployment in Figure~\ref{fig:deployment_flow} as detailed in the following.

\paragraph{Core Network Orchestration}
The 3GPP standard suggests automated mechanisms that manage dynamic deployment of VNFs, through frameworks such as NFV-MANO~\cite{ETSI_NFV_MANO}. 
During the network operation \glspl{NF} may need to be deployed, reconfigured, or terminated.
This decision is made by the VNF Manager (VNFM), a key component of the NFV-MANO framework.
Since the VNFM is responsible for deploying \glspl{NF}, a trusted deployment of NFs requires a trusted deployment of the VNFM.
The simplest way to achieve this in our threat model is to use on-premises, dedicated hardware and software stacks, following similar practices as in generations prior to 5G.
This deployment does not present any performance trade-offs, as the VNFM's responsibility to orchestrate the network does not benefit from decentralized deployment or commodity hardware and software.
In case that a decentralized VNFM deployment is desired, we propose that it is deployed as any other \gls{CN} component in our model architecture. However, this procedure must originate from operator-maintained, on-premises infrastructure, and must be initiated either manually, or by operator-managed deployment pipelines. This ensures strict remote attestation and monitoring, and provides a starting point in the chain of trust that is not exposed to the increased surface of the virtualized deployment.

\paragraph{Secure VM Configuration and Initialization}
The Secure Boot, SEV-ES, and SEN-SNP features of the host infrastructure need to be enabled in the BIOS (or cloud BIOS) before booting the VM. This configuration can be automated in a VM template that is used every time a new NF needs to be deployed by the VNFM. Before a new NF initialization, the VNFM submits the VM template request and verifies the initial remote attestation of the SVM. After validating the signed hardware and software measurements, the VNFM uses the secret key obtained from the remote attestation procedure to establish a secure communication channel with the SVM, and proceed with the NF provisioning.

\paragraph{NF Provisioning, Operation and Termination} Once the initial remote attestation is validated the VM is considered trusted. The VNFM provisions the VM with the required NF data and code through the secure communication channel. The NF operation proceeds normally, and the data in-use are protected by the TEE. The VNFM can periodically request remote attestation reports from the SVM, to validate the health of the execution environment throughout operation. This feature is supported by AMD SEV but may not be supported by alternative TEEs. In this case, the SVM would need to be re-deployed to trigger a fresh attestation report.

It is crucial that data in-flight and at rest are also protected. This can be achieved by mechanisms like secure transport and Full Disk Encryption (FDE) that are beyond the scope of our threat model. We note that these mechanisms benefit from our proposed architecture on a TEE because the secret keys that enable them are not exposed to the rest of the system during execution. 

When the NF is no longer needed, the VNFM retrieves all desired data (e.g. logs) and terminates the VM.
Since data and code on persistent storage are stored in encrypted form, their security is guaranteed by FDE mechanisms.

\subsection{Model Analysis}
In this section we provide an analysis of how our proposed model architecture reinforces trust in the presence of untrusted execution environments, and how it improves the 5G CN security against our expanded threat model. Finally, we provide important notes and considerations for the deployment of our model architecture.

\paragraph{Vertical Zero Trust}
In the previous sections we discussed how the transition to virtualized deployment and heterogeneous execution environments has created interest for ZTA in 5G. We also demonstrated the paradox that ZTA principles are vulnerable to exploits originating from an environment they are designed to protect against. Our model 5G CN architecture reinforces the basic principles of ZTA by eliminating the host infrastructure from the attack surface. This provides confidentiality and integrity to critical parts, such as digital identities and network monitoring logs, while in-use. Further, it adopts a ZTE approach by adhering to the ZTE tenets we define in~\ref{subsec:vertical_zt}:

\begin{enumerate}
    \item As per the first ZTE tenet, we treat all code, data, and computation sources, as execution resources.
    \item Our model architecture does not treat on-premises execution as trusted, and suggests confidential computing throughout the whole deployment environment. This can be relaxed up to a degree depending on the demands of every use case.
    \item According to the third ZTE tenet, trust in the execution resources is granted on a per-session basis: Our root of trust lies in the hardware and is proven via remote attestation from the TEE. Requesting a fresh remote attestation from the TEE represents a session renewal.
    \item Confidential computing extends the metrics for dynamic policies with crucial information: the observable state of the enclave, the elapsed time since the latest remote attestation, and attested properties of the software and hardware.
    \item We stress the importance of constant monitoring, which also benefits from the data confidentiality guaranteed by the TEE.
    \item Finally, we emphasize how confidential computing strictly enforces authentication and integrity mechanisms before any execution is performed, as per the sixth tenet. These mechanisms are the memory encryption and integrity protection provided by the TEE during the computation, complemented by orthogonal techniques such as FDE and TLS for secure data storage and transport.
\end{enumerate}

Our architecture, along with the assumptions derived by the ZTE tenets, provides a Zero Trust view of an execution deployment as described at the end of Section~\ref{sec:sec_model}.

\paragraph{Towards a secure 5G SBA} Our model architecture defends against the 5G SBA security threats described in Section~\ref{sec:sec_model}.

\begin{itemize}
    \item \textit{Shared memory:} The VM memory is encrypted and integrity protected by SEV, mitigating all threats arising from this exposure.
    
    \item \textit{Homogeneous software solutions:} SEV remote attestation
    provides measurements for all applications running in the VM, ensuring
    secure versions are running. We expand on this later in this section.
    
    \item \textit{Compromised Virtualization Layer:} SEV places the
    virtualization layer in the adversary model, that is, all SEV security measures protect against a compromised hypervisor.
    
    \item \textit{MANO Framework:} We propose deployment of the VNFM on operator-owned infrastructure
    or SEV encrypted VMs to mitigate the expanded attack surface.
    This presents a TCB equivalent to that of a privately-owned infrastructure, establishing
    a strong initial root of trust point in deploying NFs.
    
    \item \textit{Secure handling of sensitive data:} Techniques like secure
    transport and FDE protect data in-flight and at-rest respectively. By
    invoking these measures in SEV encrypted VMs we add extra layers against
    leaking secret keys to untrusted parts of the system.
    
\end{itemize}

\subsection{Other considerations and caveats}

\paragraph{VM TCB}
Our adversarial model considers the host software and hardware untrusted but does not provide guarantees about the software inside the VM. This is specially important when VM-based TEEs are used because of their increased TCB compared to application-based TEEs. We note that the attack vector on the NF software, OS, and libraries already existed in previous generations and are not a byproduct of the transition to the virtualized architecture. Instead, research shows that the most common cause of software exploits is human error~\cite{information_systems_society} and traditional practices such as formal verification, penetration testing, or auditing should be applied to mitigate these threats.
Nonetheless, our model provides additional layers against such threats, even when operator-owned infrastructure is used:
\begin{enumerate}
    \item Remote attestation measures all software executed in the SVM and
    guarantees that secure versions are running, i.e. versions that have
    been audited and patched against known threats.
    This simplifies the security evaluation of NFs developed by
    third party vendors which may use general purpose libraries. In case of 
    a security upgrade in a single library, the remote attestation can be
    updated to ensure the secure version is being executed, instead of
    auditing all software that may use this library.
    \item Code integrity protection by SEV ensures that external sources cannot inject
    malicious code in the SVM that would trigger potential exploits.
    Similarly, data confidentiality by SEV ensures that even if sensitive data were leaked through an exploit,
    they would not be accessible by the rest of the system.
    \item Remote attestation and constant network monitoring provide additional means to
    detect the exploit, and find and prosecute the threat agent.
\end{enumerate}
Recent advances in micro-virtualization enable efficient clusters of microVMs~\cite{firecracker_microVMs}, which are \glspl{VM} with a minimal software stack and kernel libraries, restricted only to what is absolutely necessary for the application. This largely diminishes the TCB for a VM-based TEE such as AMD SEV.

\paragraph{Side-channels} As discussed in Section~\ref{sec:sec_model}
side channel attacks are not included in the TEE threat model, and are
not handled by our proposed architecture. We consider the platform provider
responsible to provide the necessary physical security and anti-tampering
countermeasures to protect their data center servers. Application developers
should also apply the necessary security practices discussed in
Section~\ref{sec:sec_model}. This is in line with the assumption that the
platform provider is a certified, non-malicious actor and does not further
increase the attack surface of the architecture, as is the case with
operator-owned deployments of previous generations of cellular networks.
Finally, certain TEEs such as the AMD SEV provide increased protection against side-channel
attacks.

\paragraph{Deployment trade-offs}
Using the AMD SEV technology to enhance security and risk tolerance involves system design decisions that present trade-offs in the management and flexibility of the deployment:

\begin{enumerate}
    \item{Availability: AMD SEV is a proprietary technology, dependent on AMD EPYC processors. Most cloud providers offer these processors for computation purposes regardless of their AMD SEV support, and their pricing point is comparable to other data center oriented compute machines such as the Intel Xeon series of processors.
    SEV also requires compatibility support by the hypervisor and guest OS, but this is
    increasingly integrated in the latest versions of popular kernel and OS releases.
    Finally, AMD SEV is rapidly deployed on cloud provider platforms in
    beta or general access instances at a similar price-point as a non-confidential
    machine of the same computational power. We provide a cost evaluation of our deployment in Section~\ref{sec:results}.
    Overall, SEV has been deployed since 2016 and is an increasingly available and affordable solution.
    As the VM-based TEE ecosystem evolves, with more manufacturers joining the market and cloud providers offering more confidential computing solutions, the availability of those technologies is becoming a minor concern.}

    \item{VM features: The design of SEV explicitly removes the hypervisor from
    the attack surface by encrypting the SVM memory. Naturally, the 
    hypervisor loses the ability to perform certain functionalities like
    VM memory snapshots, hot-add devices, suspend \& resume, hot clones,
    live migration, PCI-passthrough and vMotion.
    It is likely that many of these limitations will be removed by future
    hardware, firmware, and software features~\cite{suse_sev_doc}, and AMD
    SEV already added support for live migration of encrypted VMs. Additionally,
    with clever application design a service can be resilient to the absence of many of these features. Specifically in the context of the 5G CN, the NFs are designed to provide a continuous and dynamic service to connected devices
    that does not benefit from VM memory snapshots, hot clones, and suspend \& resume operations:
    by the time a NF is resumed, the status
    of the rest of the network most likely changed and must be reassessed.
    Hot-add devices and PCI-passthrough are not frequent or critical in decentralized deployments
    of a Service-Based Architecture.
    Finally, migration procedures are ideally handled internally by the 5G CN
    so that connected NFs and served UEs are reassigned to
    other network components before an NF is migrated to another host.}
\end{enumerate}

\section{Experimental evaluation} \label{sec:results}
\noindent In order to assess the viability of our model architecture for B5G networks we set up a testbed using 5G open-source software, and perform experiments to determine the performance overhead of 5GC-specific operations.
We deploy an end-to-end 5G cellular network testbed, including the CN, RAN, and UE, on both cloud and private infrastructure. Cloud platforms represent a more realistic 5G SBA deployment, whereas our private infrastructure benefits from a controlled environment, and provides more accurate performance benchmarks, by eliminating factors like multi-tenancy effects observed in the cloud~\cite{cloud_multitenancy_profiling}.
We use the Open5GS~\cite{Open5GS} framework for the CN, UERANSIM~\cite{ueransim} for RAN and UE simulation, and Microsoft Azure Cloud and Google Cloud
as cloud providers. The Azure cloud VMs support AMD SEV-SNP in general access on AMD EPYC 7763 3rd gen. (Milan) processors,
and the Google cloud VMs support AMD SEV-ES in a beta release on AMD EPYC 7B12 2nd gen. (Rome) processors.
On our private infrastructure, we deploy VMs on a server equipped with an AMD EPYC 7443P 24-Core CPU and SEV-SNP support. All VMs (both cloud, and privately hosted) are allocated 4-core CPUs and 16GB of RAM.
Each setup consists of 6 VMs that host the 5G Control Plane, 5G User Plane, RAN, and UE, with the first two being deployed on both SEV-enabled
and SEV-disabled VMs. By isolating the computation load of the RAN and UE and separating the control plane and user plane we are able to accurately
measure the performance difference of confidential machines in different situations such as low-traffic control signals and database operations, and high-traffic user data. In order to simulate UE traffic we generate \textit{iPerf3} traffic of varying throughput from every UE, to an external Data Network (i.e. Internet) through the Open5GS UPF.

We benchmark the performance of 5G-specific functionalities: a) creating the UE database in the 5G core, b) registering a NF to the NRF, and c) connecting new UEs to the network, by measuring the average elapsed time in every case. All experiments were conducted both on cloud and our private infrastructure to provide a view of the impact of AMD SEV in different deployment scenarios.

\begin{figure}[t]
     \includegraphics[width=\columnwidth]{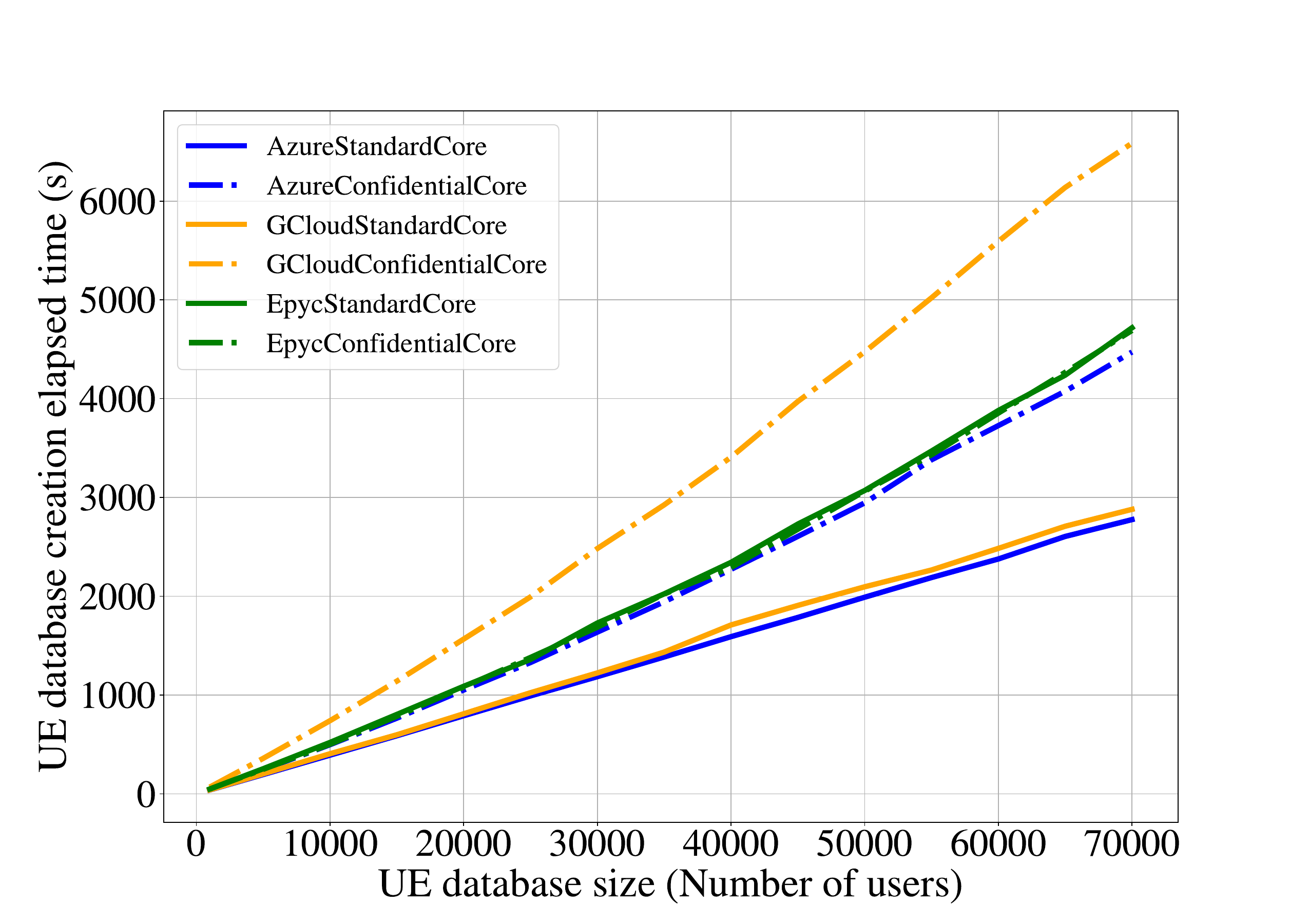}
     \caption{Average time elapsed for UE database creation in the 5G Core.}
     \label{fig:create_db}
\end{figure}

\begin{figure}[t]
     \vfill
     \includegraphics[width=\columnwidth]{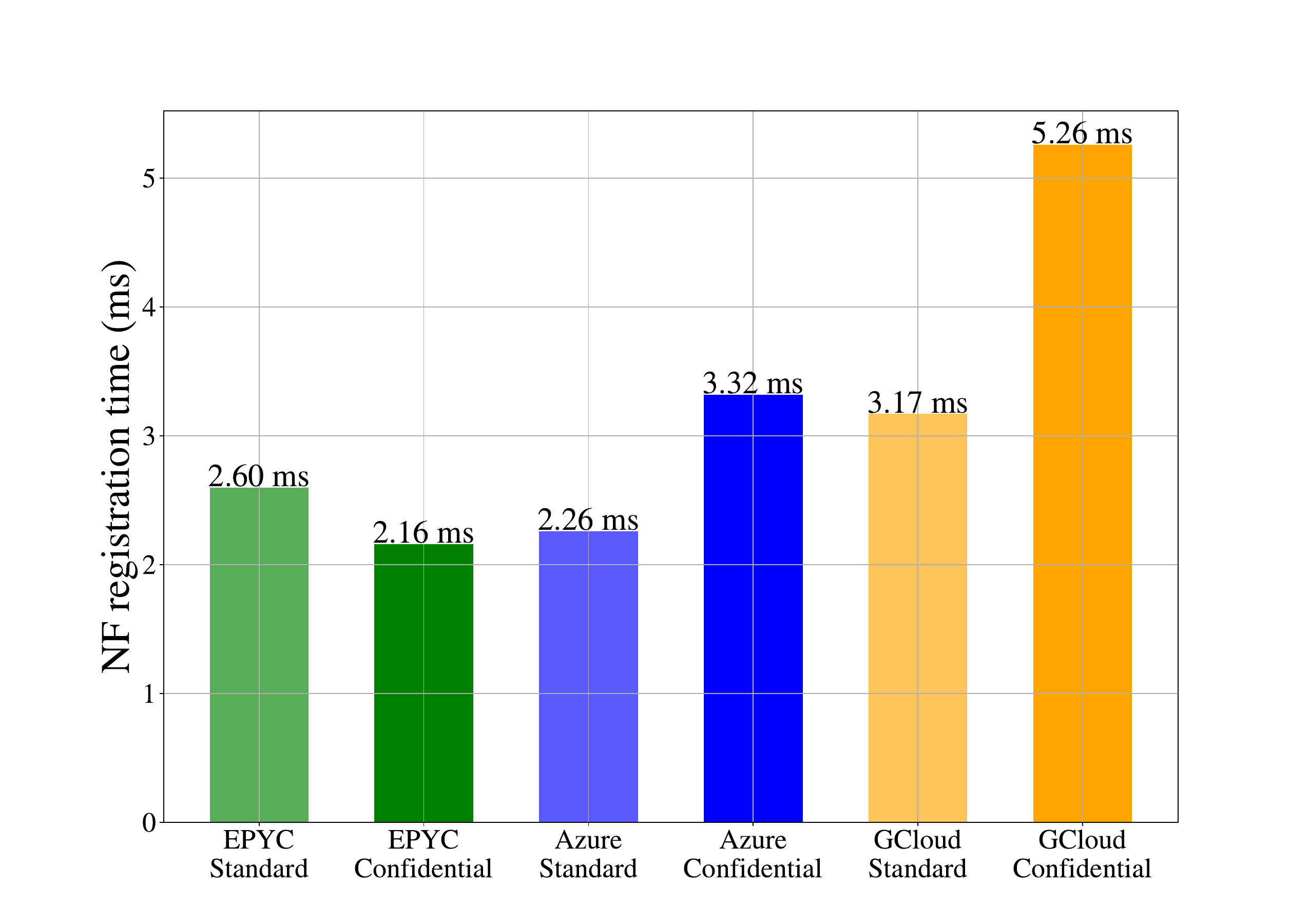}
     \caption{Average time elapsed for NF registration to the NRF.}
     \label{fig:nf_reg}
\end{figure}

\begin{figure}[t]
     \vfill
     \includegraphics[width=\columnwidth]{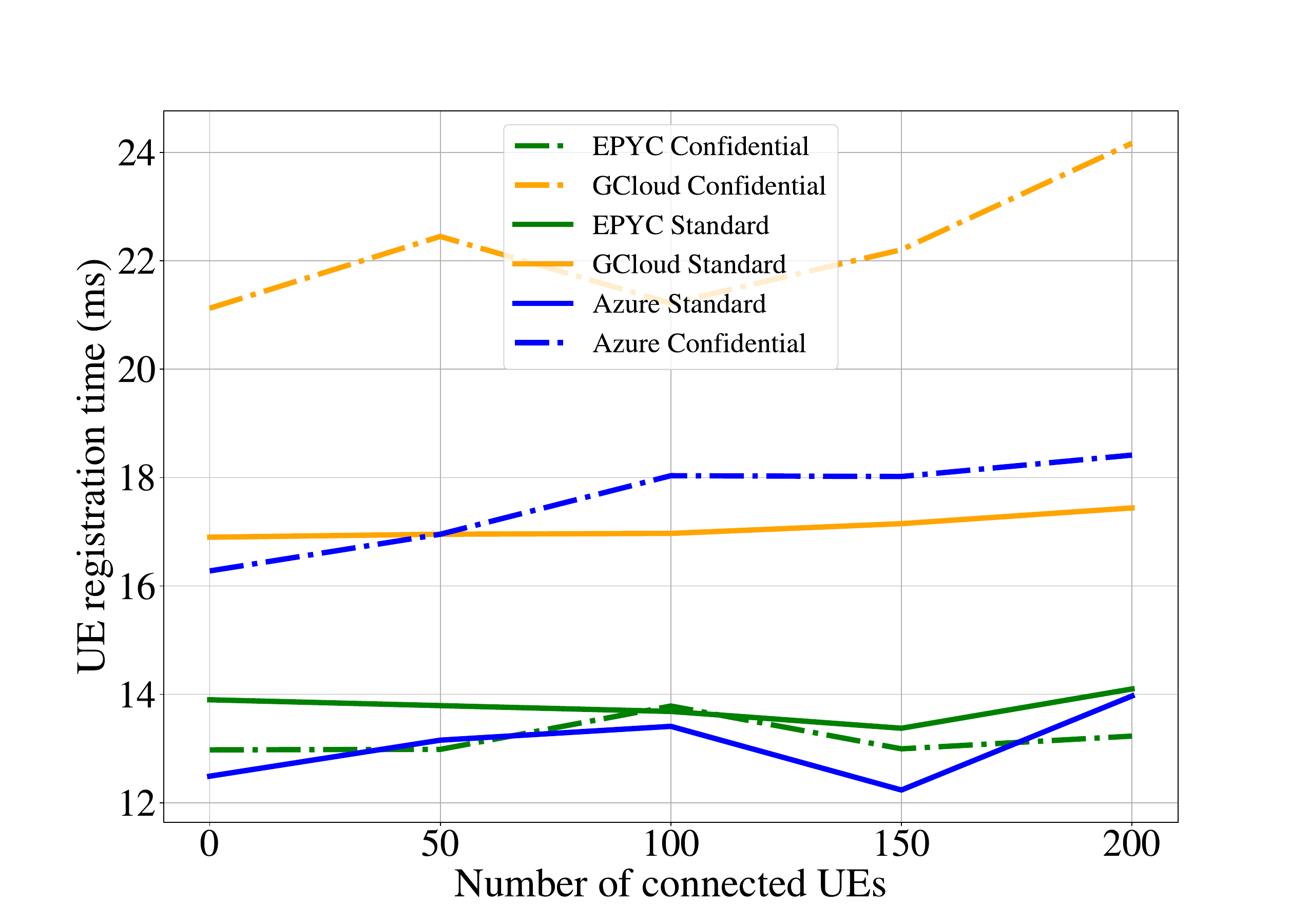}
     \caption{Average time elapsed for UE registration to the network.}
     \label{fig:ue_reg}
\end{figure}

We first set up the 5G CN in confidential and standard VMs, without connecting the RAN or UE. Figure~\ref{fig:create_db} depicts the elapsed time for the creation of UE databases of variable sizes in the CN, averaged over 100 executions.
We observe a small overhead for the cloud confidential VMs which is not present in our private server. We attribute this to load balancing and resource sharing effects in cloud VMs
that are not present in the controlled environment of our private server.
In Figure~\ref{fig:nf_reg} we show the averaged elapsed time over 100 executions for the AMF registration to the NRF,
and observe a similar pattern. Finally, Figure~\ref{fig:ue_reg} shows the average elapsed time for the registration and connectivity establishment of a UE with the CN. The use of confidential VMs increases the elapsed time by at most 7ms compared to the standard VM, which is an acceptable overhead for the UE registration functionality, and much smaller than the limitations defined in the 5G standard. In Figures~\ref{fig:nf_reg},~\ref{fig:ue_reg} the performance of our private server in confidential mode is slightly better than in standard mode. The difference is within fractions of a millisecond and falls into margins of error due to CPU scheduling and logging timing, and we consider them almost equivalent.
All benchmarks show that the system scales linearly with the workload increase, making it viable for large-scale deployments such as the 5G CN. We find that our measurements align with the general-purpose benchmarks provided by AMD, which show a 2-8\% performance overhead when using AMD SEV on Azure Confidential Cloud.

Finally, we discuss the monetary cost overhead that comes with the AMD SEV deployment, using publicly available pricing data by manufacturers and cloud providers at the time of writing. AMD SEV is a feature incorporated in the EPYC series of AMD data center oriented processors. EPYC processors have been widely leveraged for high performance computing. Their cost is comparable to the equivalent Intel Xeon processor, at 90\% for cloud deployment.
Enabling the extra feature of AMD SEV on an EPYC processor comes \textit{almost} for free in terms of monetary cost.
In the case of private servers, AMD SEV requires a single TPM module which costs less than 2\% of the cost of the least powerful 3rd gen. EPYC CPU (AMD EPYC 7443P 24-Core Processor). In the case of cloud provided VMs, a confidential CPU core has a cost overhead of less than 10\% compared to its non-confidential counterpart, being around \$0.01 per core per hour more expensive.

\section{Conclusion}\label{sec:conclusion}
\noindent In this work, we provide a study on design, practical issues, and trade-offs for establishing trust in the Beyond-5G cellular Core Network using Trusted Execution Environments.
This analysis will help the community to better understand the augmented attack surface in B5G cellular networks and the importance of zero trust principles.
We analyze the security implications of the 5G Service Based Architecture with respect to the execution environment.
Specifically, we extrapolate the Zero Trust principles already supported by 5GC to the execution layer,
and define Zero Trust Execution, a set of principles that formalize trust establishment in the execution stack.
We propose the use of Trusted Execution Environments to reinforce those principles. We propose a 5GC System Architecture leveraging the VM-based AMD SEV as our TEE platform, because of its security guarantees, efficient deployment, and maturity. We set up a testbed to measure the impact of AMD SEV on the 5G CN operation, and benchmark the elapsed time of functionalities such as UE database creation, NF registration and UE registration. We show that the additional latency is minimal, and well within the requirements of the 5G standard. We conclude that the 5G CN confidential deployment comes with minimal impact to the overall performance, and provides valuable security guarantees. This, combined with the affordable monetary cost overhead, makes our proposed 5G Confidential Core Network a very viable addition to the security and trust mechanisms in the 5G Core Network.


\end{document}